# Relaxor ferroelectricity and colossal magnetocapacitive coupling in ferromagnetic CdCr$_2$S$_4$

J. Hemberger[1], P. Lunkenheimer[1], R. Fichtl[1], H.-A. Krug von Nidda[1], V. Tsurkan[1,2] & A. Loidl[1]

[1]*Experimental Physics V, Centre for Electronic Correlations and Magnetism, University of Augsburg, 86159 Augsburg, Germany*
[2]*Institute for Applied Physics, Academy of Sciences of Moldova, MD-2028, Chisinau, R. Moldova*

**Multiferroic materials, which reveal magnetic and electric order, are in the focus of recent solid state research[1-4]. Especially the simultaneous occurrence of ferroelectricity and ferromagnetism, combined with an intimate coupling of magnetization and polarization via magneto-capacitive effects, could pave the way for a new generation of electronic devices. Here we present measurements on a simple cubic spinel with unusual properties: It shows ferromagnetic order and simultaneously relaxor ferroelectricity, i.e. a ferroelectric cluster state, reached by a smeared-out phase transition, both with sizable ordering temperatures and moments. Close to the ferromagnetic ordering temperature the magneto-capacitive coupling, characterized by a variation of the dielectric constant in an external magnetic field, reaches colossal values of nearly 500%. We attribute the relaxor properties to geometric frustration, which is well known for magnetic moments, but here is found to impede long-range order of the structural degrees of freedom.**

The coexistence of ferroelectricity and ferromagnetism would constitute a mile stone for modern electronics and functionalised materials. The most appealing applications are new types of storage media using both magnetic *and* electric polarization and the possibility of electrically reading/writing magnetic memory devices (and vice versa). However, it is clear now that ferroelectric ferromagnets are rare[5,6] and mostly exhibit rather weak ferromagnetism. Spinel compounds are an important class of materials and their electronic properties are in the focus of research since the famous work of Verwey on magnetite[7]. Recent reports on geometrical frustration of the spin and orbital degrees of freedom[8], and the observation of an orbital-glass state[9] in sulpho spinels, demonstrate the rich and complex physics, characteristic of these compounds. Here we report on another interesting experimental observation in a spinel system: Relaxor ferroelectricity in ferromagnetic CdCr$_2$S$_4$ and the occurrence of colossal magneto-capacitive effects.

CdCr$_2$S$_4$ crystallizes in the normal cubic spinel structure (space group Fd3m, $a = 1.024$ nm), with Cr$^{3+}$ octahedrally surrounded by sulphur ions, yielding a half-filled lower t$_{2g}$ triplet with a spin $S = 3/2$. Ferromagnetism in CdCr$_2$S$_4$ is well known[10], but early experimental observations of a number of mysterious features have fallen into oblivion: For example, reports of an anomalous expansion coefficient at low temperatures[11,12], an unexpected concomitant broadening of the diffraction lines[11], a strong blue shift of the absorption edge on passing the ferromagnetic phase transition[13], the observation of anomalously large phonon shifts and damping effects close to $T_c$[14] and the observation of large magneto-resistance effects[15].

Figure 1a shows the inverse magnetic susceptibility $c^{-1}$ and the low-temperature magnetization $M$. The straight line indicates a fit to the paramagnetic susceptibility, which results in a Curie-Weiss temperature of 155 K and a paramagnetic moment of 3.88 $\mu_B$, close to the theoretically expected value of 3.87 $\mu_B$ per Cr$^{3+}$. The deviations when approaching $T_c$ arise from ferromagnetic fluctuations. $M(T)$ strongly increases at $T_c$, the further increase just below $T_c$ being interrupted by demagnetization effects. The inset in Fig. 1a shows the ferromagnetic hysteresis revealing soft-magnetic behaviour and a saturation magnetization of 3 $\mu_B$ per Cr$^{3+}$. Hence, CdCr$_2$S$_4$ behaves like a typical soft ferromagnet.

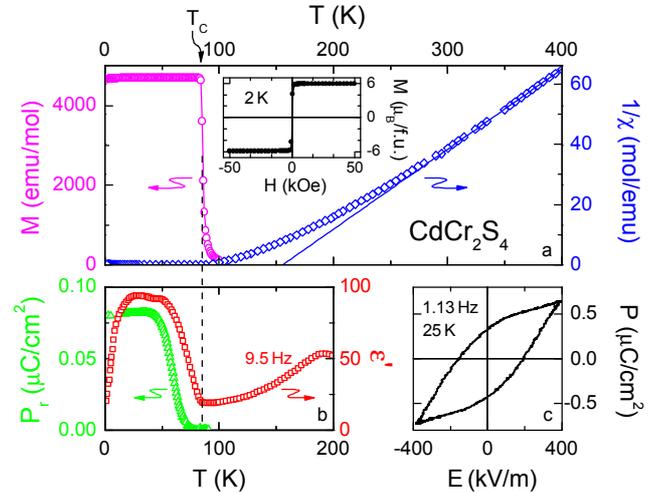

**Figure 1** Magnetic and dielectric characterization of CdCr$_2$S$_4$. **a,** Right scale: Inverse magnetic susceptibility *vs.* temperature measured at 100 Oe. The solid line indicates a fit ($T > 300$ K) using a Curie-Weiss law, with a Curie-Weiss temperature of 155 K. Left scale: Magnetization *vs.* temperature measured at 100 Oe. Inset: Ferromagnetic hysteresis at 2 K, indicating a saturated moment of 6 $m_B$ per formula unit. **b,** Right scale: Dielectric constant *vs.* temperature measured at a frequency of 9.5 Hz. Left scale: Thermo-remanent polarization *vs.* temperature measured after cooling in an electric field of 50 kV/m. **c,** Polarization *vs.* electric field showing a ferroelectric hysteresis.

Figure 1b shows the temperature dependence of the dielectric constant, measured at 9.5 Hz (right scale). $e'(T)$ exhibits a steep rise below $T_c$, obviously driven by the onset of ferromagnetic order. Strong magneto-electric coupling effects must be responsible for this behaviour. At lower temperatures, $e'(T)$ decreases again, leading to a maximum reminiscent of ferroelectric behaviour. However, its rather large width and the only moderately high maximum values of $e' \approx 100$ indicate a transition into a polar state of diffusive character. Indeed, a finite polarization, a direct proof of ferroelectricity, arises well below $T_c$ (Fig. 1b, left scale). Final evidence is provided in Fig. 1c, which shows a polarization cycle using electric fields up to 400 kV/m exhibiting a clear ferroelectric hysteresis. But the relatively slim loop indicates non-canonical ferroelectric behaviour[16].

Evidence for the actual character of this ferroelectric state is revealed from the temperature dependence of $e'$, shown in Fig. 2a for a series of measuring frequencies. A very broad peak appears between 150 and 250 K, shifting to lower temperatures and increasing in amplitude with decreasing frequency. It exhibits the characteristic behaviour of a relaxor ferroelectric, the typical strong dispersion effects often being ascribed to the freezing-in of ferroelectric clusters[16,17] In contrast to conventional ferroelectrics, relaxor ferroelectrics are characterized by a diffuse phase transition, extending over a finite temperature range, by strong relaxational dispersion effects in dielectric constant and loss and by a macroscopic polarization showing up only well below the transition temperature[16-18]. Most relaxor ferroelectrics known so far are perovskite-related materials where long-range polar order is suppressed by substitutional disorder[17]. However, the relaxor properties in CdCr$_2$S$_4$ occur in a pure compound without any disorder. By comparison to measurements with silver-paint contacts

(not shown), the increase at temperatures above the peak is found to be due to contact effects[19], whereas the peak itself is clearly an intrinsic property of $CdCr_2S_4$. The high-temperature flank of the relaxor peaks can roughly be described by a Curie-Weiss law with a characteristic temperature of 135 K (dashed line). In relaxors, often deviations from a Curie-Weiss law are observed; it gives, however, a rough estimate of a quasistatic freezing temperature. Notably, significant spontaneous polarization sets in only well below 135 K, a behaviour often found in relaxor ferroelectrics[17]. Further evidence for typical relaxor behaviour arises from the frequency dependence of the dielectric loss $e''$. The inset of Fig. 2a shows $e''(n)$ for a series of temperatures, obtained after subtraction of a conductivity contribution. Broad peak maxima are revealed, typical for relaxational behaviour. The peak frequency decreases when the temperature is lowered, indicating the freezing-in of polar moments. In $CdCr_2S_4$ the half width of the loss peak at 151 K amounts nearly three decades in frequency, significantly broader than the 1.14 decades expected for a Debye relaxation, indicating a broad distribution of relaxation times. However, in most canonical relaxors even broader distributions are found[17,18] and we cannot exclude that contributions from hopping charge carriers (e.g. small polarons trapped in defects[20]) play a role in the appearance of the observed relaxation feature.

low frequencies becomes increasingly suppressed for higher frequencies. This can be explained assuming that the polar moments, which at about 100 K are frozen-in at the time-scale of the experiment, speed up again below $T_c$. Obviously, the increasing ferromagnetic correlations accelerate the mean polar relaxation rates. This leads to a remelting of the frozen-in relaxor state at $T < T_c$, resulting in strongly enhanced dielectric constants. Only when the temperature is further lowered (below about 20 K), finally "temperature wins" and $e'$ becomes reduced due to a complete freezing-in of the polar dynamics. Most probably, exchangestriction softens the lattice, reduces the energy barriers against dipolar reorientation and enhances the mean relaxation rates. Exchangestriction was also utilized to explain the enormous blue shift in $CdCr_2S_4$[21]. On the other hand, a contribution from hopping-type charge transport to the anomaly at $T_c$ via a variation of the ac conductivity, affecting also $e'$, cannot be excluded, especially as it is well known that $CdCr_2S_4$ exhibits sizeable magneto-resistance effects[15]. In any case, overall the magneto-capacitive coupling in $CdCr_2S_4$ is different to all mechanisms observed so far and purely dynamic in origin.

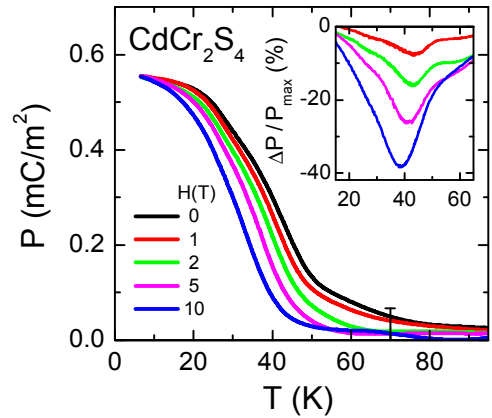

**Figure 3** Thermo-remanent polarization *vs*. temperature measured after cooling in an electric field of 150 kV/m in various external magnetic fields, directed perpendicular to the electric field. The error bar indicates an 80% confidence error caused by conductivity contributions. The inset shows the magnetic-field dependent variation of the polarization, $DP=P(H)-P(0T)$, related to the maximum polarization of about 0.55 mC/m$^2$.

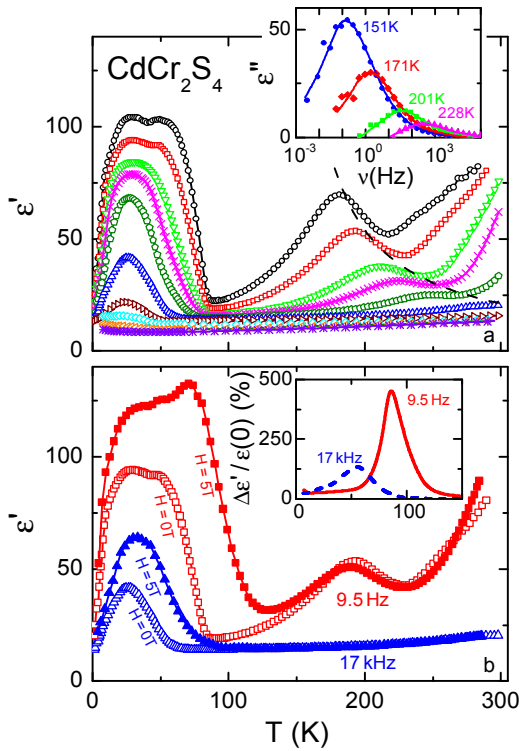

**Figure 2** Magneto-capacitive behaviour of $CdCr_2S_4$. **a**, Temperature dependence of $e'$ at various frequencies (from top to bottom: 3 Hz, 9.5 Hz, 53 Hz, 170 Hz, 950 Hz, 17 kHz, 950 kHz, 12 MHz, 83 MHz, 3 GHz). The dashed line indicates the static dielectric susceptibility following a Curie-Weiss like law with a characteristic temperature of 135 K. The inset shows $e''(n)$ for various temperatures above 150 K (the lines guide the eye). **b**, Dielectric constant *vs*. temperature at 9.5 Hz and 17 kHz, measured at zero field and in an external magnetic field of 5 T, directed perpendicular to the electric field. The inset provides a measure of the magneto-capacitive effects, with $De' = e'(5T) - e'(0T)$, for the two frequencies shown in the main frame.

Interestingly, a second relaxation-derived dispersion with a concomitant strong increase of $e'$ appears at temperatures below the onset of ferromagnetic order (Fig. 2a). A double peak appearing at

Figure 2b demonstrates that the magneto-capacitive coupling indeed is extremely strong and can be termed colossal. It shows the temperature dependence of the dielectric constant at 9.5 Hz and 17 kHz as measured in zero magnetic field and in 5 T. The relaxation dynamics at $T > T_c$ remains essentially unaffected by the field. However, strong magneto-capacitive effects are observed close to $T_c$. The magneto-capacitance as derived from the difference of the dielectric constant in zero and non-zero fields is shown in the inset of Fig. 2b. At low measuring frequencies, the maximum effects occurring close to $T_c$ reach 450 % at 9.5 Hz. For possible applications of this phenomenon, certainly some hurdles have to be overcome, e.g., by doping or searching for related multiferroic spinel systems with higher transition temperatures.

Finally, Fig. 3 shows the thermo-remanent polarization for various magnetic fields, demonstrating a strong coupling of both quantities. Increasing external magnetic fields suppress the polar state, in marked contrast to the findings in other magneto-capacitive materials[1-3]. This is fully consistent with our assumption that the onset of ferromagnetic order enhances the reorientational mobility of the polar entities. Increasing magnetic fields drive the system towards ferromagnetism and thus the macroscopic polarisation breaks down already at lower temperatures for higher fields (Fig. 3).

The polarization shift increases with magnetic field (inset of Fig. 3), documenting the strong competition between ferroelectricity and ferromagnetism. The magnetic-field dependence shown in Fig. 2b and 3 was measured with E⊥H. Measurements with E∥H led to identical results.

Finally, two more mysteries remain to be resolved: What is the nature of the ordered dipolar moment in $CdCr_2S_4$ and how could relaxor ferroelectricity occur without disorder-derived frustration? It is clear that the polar ordering must be displacive in nature, driven by soft-mode behaviour as in ferroelectric perovskites. This notion is consistent with the mentioned structural softening deduced in ref. 11. Indeed the observation of polar displacements of the octahedrally coordinated ions has been reported for a number of spinels[22]. Guided by these results, we suggest that ferroelectricity in $CdCr_2S_4$ results from an off-centre position of the $Cr^{3+}$-ions. It may well be that spinel compounds will turn out to be a new class of semiconducting ferroelectrics.

In relaxor ferroelectrics, the freezing of polar moments is driven by frustrated interactions related to disorder. For magnetic systems, frustration can also arise in ordered solids from geometrical constraints alone[23]. Ising spins with nearest-neighbour antiferromagnetic interactions on a triangular lattice are a standard example. The possibility that geometric frustration might also exist in non-magnetic systems has been suggested by Ramirez[24], exemplified by $ZrW_2O_8$ revealing negative thermal expansion[25]. This effect has been explained with a concept of frustrated soft-mode behaviour[26]. $CdCr_2S_4$ also shows negative thermal expansion and reveals a strong broadening of diffraction lines below 120 K[11,12]. The broadening of diffraction lines on cooling provides clear evidence for the loss of true long-range order and indicates that local order may substantially differ from global symmetry. Similar effects have been observed in the orientational glass KBr:KCN[27]. In orientational glasses[28], disorder-derived frustration drives the glassy freezing. In contrast, the suggested polar freezing in $CdCr_2S_4$ could be governed by geometrical constraints.

The coexistence of ferromagnetism and proper ferroelectricity with reasonable ordering temperatures and sizable saturated moments, in addition to colossal magneto-capacitive effects and strong coupling of ferroelectric polarization to external magnetic fields, are the most interesting aspects of the behaviour of $CdCr_2S_4$. Compared to recent reports on other multiferroics[1-4], this spinel system is unique: The perovskite-derived manganites treated in refs. 1-3 are improper ferroelectrics with polar order induced by modulated spin structures, accompanied by lattice modulations. In contrast, the hexagonal manganite of ref. 4 is a proper ferroelectric with high ordering temperature ($T_c = 875$ K) and much lower antiferromagnetic ordering ($T_N = 75$ K). Ferromagnetism is induced by applied electric fields. In $CdCr_2S_4$ ferromagnetism and ferroelectricity to first order develop independently, driven by electronic superexchange and soft-mode behaviour, respectively. Both phenomena are linked to the $Cr^{3+}$ ions and strongly compete yielding colossal magnetoelectric effects. In $CdCr_2S_4$, magnetization and polarization, as well as dipolar and magnetic susceptibilities are compatible with a global cubic symmetry. All quantities are independent of the directions of the magnetic with respect to the electric field, indicating that $CdCr_2S_4$ belongs to a new class of multiferroics. Despite these differences, frustration seems to play a fundamental role in all these multiferroic compounds: It is frustration of the electronic degrees of freedom in the manganites and geometrical frustration of the lattice degrees of freedom in the spinel.

## Methods

Single crystals of $CdCr_2S_4$ were grown by chemical transport-reaction using chlorine as transport agent. Starting material was the ternary compound obtained by standard ceramic techniques. The single crystals were routinely characterized by powder and single crystal x-ray diffraction and by wave-length sensitive electron-probe microanalysis. We found an almost ideal stoichiometry and we can exclude any site inversion between the cations. To check for the robustness of the results, we performed measurements on several crystals from different batches. All samples revealed ferromagnetic and ferroelectric ordering temperatures well within the experimental uncertainties. Magnetic susceptibility and magnetization were measured using a commercial SQUID. For the dielectric measurements of the present work, sputtered gold-contacts were applied to the plate-like samples forming a parallel-plate capacitor. To check for possible contributions from electrode polarization, the same sample was also measured with silver-paint contacts. The dielectric constant and loss were measured over a broad frequency range of nine decades (3 Hz < ν < 3 GHz). A frequency-response analyzer (Novocontrol α-analyzer) was used for frequencies $n < 1$ MHz and a reflectometric technique employing an impedance analyzer (Agilent E4291A) at $n > 1$ MHz[29]. The temperature-dependent polarization was determined via the pyro charge measured with an electrometer (Keithley 617). The same device was used as impedance transformer within a Sawyer-Tower circuit, read out by a two-channel digital oscilloscope (Tektronik TDS210) and stimulated via a high-voltage amplifier (Trek 609). For cooling to temperatures down to 1.4 K, a conventional $^4$He bath-cryostat (Cryovac Konti-IT) was employed; additional magnetic-field dependent measurements up to 10 T were performed in an Oxford cryomagnet. The magnetic anisotropy was checked by angular-dependent SQUID and electron-spin-resonance measurements. We found almost isotropic behaviour in the paramagnetic phase, and a slight cubic anisotropy in the magnetically ordered phase. All dielectric measurements were performed also measuring along different crystallographic directions and with E∥H and E⊥H, revealing identical results within experimental resolution.

### Acknowledgements
This work was partly supported by the Deutsche Forschungsgemeinschaft via the Sonderforschungsbereich 484 and partly by the BMBF via VDI/EKM.

**Correspondence** and requests for materials should be addressed to P.L. (peter.lunkenheimer@physik.uni-augsburg.de).

[1] Kimura, T. *et al.* Magnetic control of ferroelectric polarization. *Nature* **426**, 55-58 (2003).
[2] Goto, T., Kimura, T., Lawes, G., Ramirez, A. P. & Tokura, Y. Ferroelectricity and giant magnetocapacitance in perovskite rare-earth manganites. *Phys. Rev. Lett.* **92**, 257201-1-257201-4 (2004).
[3] Hur, N. *et al.*, Electric polarization in a multiferroic material induced by magnetic fields. *Nature* **429**, 392-395 (2004).
[4] Lottermoser, Th. *et al.* Magnetic phase control by an electric field. *Nature* **430**, 541-544 (2004).
[5] Hill, N. A. Why are there so few magnetic ferroelectrics. *J. Chem. Phys. B* **104**, 6694-6709 (2000).
[6] Smolenskii, G. A. & Chupis, I. E. Ferroelectromagnets. *Sov. Phys. Usp.* **25**, 475-493 (1983).
[7] Verwey, J. E. W. Electronic conduction of magnetite ($Fe_3O_4$) and its transition point at low temperature. *Nature* **144**, 327-328 (1939).
[8] Fritsch, V. *et al.*, Spin and orbital frustration in $MnSc_2S_4$ and $FeSc_2S_4$. *Phys. Rev. Lett.* **92**, 116401-1-116401-4 (2004).
[9] Fichtl, R., Tsurkan, V., Lunkenheimer, P., Hemberger, J., Fritsch, V, Krug von Nidda, H.-A., Scheidt, E.-W. & Loidl, A. Orbital freezing and orbital glass state in $FeCr_2S_4$. *Phys. Rev. Lett.* **94**, 027601-1-027601-4 (2005).
[10] Baltzer, P. K., Lehmann, H. W. & Robbins, M. Insulating ferromagnetic spinels. *Phys. Rev. Lett.* **15**, 493-495 (1965).
[11] Göbel, H. Local lattice distortions in chromium chalcogenide spinels at low temperatures. *J. Magn. Magn. Mater.* **3**, 143-146 (1976).
[12] Martin, G. W., Kellog, A. T., White, R. L. & White, R. M. Exchangestriction in $CdCr_2S_4$ and $CdCr_2Se_4$. *J. Appl. Phys.* **40**, 1015-1016 (1969).

[13] Harbeke, G. & Pinch, H. Magnetoabsorption in single-crystal semiconducting ferromagnetic spinels. *Phys. Rev. Lett.* **17**, 1090-1092 (1966).

[14] Wakamura, K. & Arai, T. Effect of magnetic ordering on phonon parameters for infrared active modes in ferromagnetic spinel $CdCr_2S_4$. *J. Appl. Phys.* **63**, 5824-5829 (1988).

[15] Lehmann, H. W. & Robbins, M. Electrical transport properties of the insulating ferromagnetic spinels $CdCr_2S_4$ and $CdCr_2Se_4$. *J. Appl. Phys.* **37**, 1389-90 (1966).

[16] Samara, G.A. The relaxation properties of compositionally disordered $ABO_3$ perovskites. *J. Phys. Cond. Matt.* **15**, R367-R411 (2003).

[17] Cross, L. E. Relaxor ferroelectrics. *Ferroelectrics* **76**, 241-267 (1987).

[18] Kamba, S. *et al.* Dielectric dispersion of relaxor PLZT ceramics in the frequency range 20 Hz-100 THz. *J. Phys. Cond. Matt.* **12** 497–519 (2000).

[19] Lunkenheimer, P. *et al.* Origin of apparent colossal dielectric constants. *Phys. Rev. B* **66**, 052105-1052105-4 (2002).

[20] Austin, I. G. & Mott, N. F. Polarons in crystalline and non-crystalline materials, *Adv. Phys.* **18**, 41-103 (1969).

[21] Callen, E. Optical Absorption Edge of Magnetic Semiconductors. *Phys. Rev. Lett.* **20**, 1045-1048 (1968).

[22] Grimes, N. W. Off-centre ions in compounds with spinel structure. *Phil. Mag.* **26**, 1217-1226 (1972).

[23] Ramirez, A. P. Geometrical frustration, in *Handbook of Magnetic Materials* Vol. 13 (ed. Buschow, K. H. J.) 423-520 (Elsevier, Amsterdam, 2001).

[24] Ramirez, A. P. Magic moments. *Nature* **421**, 483 (2003).

[25] Mary, T., Evans, J. S. O., Vogt, T. & Sleight, A. W. Negative thermal expansion from 0.3 to 1050 K in $ZrW_2O_8$. *Science* **272**, 90-92 (1996).

[26] Cao, D., Bridges, F., Kowach, G. R. & Ramirez, A. P. Frustrated soft modes and negative thermal expansion in $ZrW_2O_8$. *Phys. Rev. Lett.* **89**, 215902-1-215902-4 (2002).

[27] Knorr, K. & Loidl A. Anomalous diffraction profiles of Alkali-Halide-Alkali-Cyanide mixed crystals. *Phys. Rev. Lett.* **57**, 460-462 (1986).

[28] Höchli, U. T., Knorr, K & Loidl, A., Orientational glasses. *Adv. Phys.* **39**, 405-615 (1990).

[29] Lunkenheimer, P., Schneider, U., Brand, R. & Loidl, A. Glassy dynamics. *Contemp. Physics* **41**, 15-36 (2000).